# Quasielastic neutron scattering by concentrated aqueous “single-nano buckydiamond” dispersion

**N.M. Blagoveshchenskii, [1] A.G. Novikov, [1*] E.Osawa, [2] N.N. Rozhkova [3]**

[1] *SSC RF – Institute of Physics and Power Engineering, Obninsk, Kaluga Reg., Russia*
[2] *NanoCarbon Research Institute, AREC, Faculty of Textile Science and Technology, Shinshu University, Ueda, Nagano, 386-8567 Japan*
[3] *Institute of Geology, KRC RAS, Petrozavodsk, Karelia, Russia*
**corresponding author: novikov@ippe.ru*

**Abstract**

The diffusion characteristics of stable aqueous dispersion of “single-nano buckydiamond” (hereinafter referred to as dispersion water DW, concentration of SNBD ~ 80 mg/ml, average particle size ~ 8 nm, further shortly ND) and pure water (hereinafter referred to as bulk water, BW) as the reference system were investigated by quasielastic neutron scattering (QENS).
The difference of BW and DW spectra were analyzed by two ways. In the first one DW was described as the two-component system, consisted of BW and a small additive having presumably the lorentzian form. This additive is called as hydration water (HW), presumably located in the nearest surrounding of ND’s. The second approach included the direct subtraction of DW and BW spectra, properly normalized in the regions of their wings, where the influence of HB can be neglected. The relative fractions of HB, extracted for both cases, in the limits of errors are near each other and equal about 3 %.
HW was found to have diffusive mobility which is retarded as compared to BW due to the weakening of the rotational component of diffusion and the increased residence lifetime of molecules.

## I. INTRODUCTION

The distinctive properties of substances in nanostate and the promising prospects of their practical application have provided an impetus to the study of the properties of these substances and methods for their production. One field of such an active research is the study of carbon nanoparticles (CNP: fullerene, carbon nanotubes, nanodiamonds, shungites, etc.) [1]. It appears that materials based on CNP offer several advantages, in particular, increased strength and durability, due to their highly developed surface. They also accelerate chemical reactions. At the same time, the highly developed surface of CNPs results in their excess surface energy, which leads to spontaneous aggregation - a considerable barrier on the way to their practical application.

There are a number of methods to stabilize CNPs in particular size and to prevent their spontaneous aggregation. An essential method is to disperse CNPs in water to take advantage of arising hydration effects. The study of hydration leads to better understanding of the aggregation mechanisms, and casts light on the roles of water as a dispersion medium and a regulator of the molecular interactions in CNPs. The stability of a aqueous dispersion system and the kinetics of aggregation processes should depend on the structural-dynamic characteristics of the CNP hydration shell that, in turn, are connected with the properties of its own hydrogen-bonds network and the pattern of "carbon - water" interaction.

Some time ago one of us succeeded [2] in preparing aqueous dispersions of the primary particles of detonation nanodiamonds to fairly high concentrations, which enabled the present study to be carried out.

## II. EXPERIMENTAL

The scattering experiments were performed on an inelastic neutron scattering spectrometer DIN-2PI [3], attached to the IBR-2 pulsed reactor installed in the Frank Laboratory of Neutron Physics (JINR, Dubna, Russia). The initial neutron energy $E_o = 3\ meV$ (resolution $\Delta E_o \sim 0.14\ meV$) with the neutron wave number transfer $Q < 2.2\ \AA^{-1}$ was used in the experiment. The measurements were carried out on the aqueous dispersion of nanodiamond particles (vide infra) and bulk water (BW) as reference system. The liquids under study were placed in an aluminum container, 0.5 mm thick, 120 mm in height and 100 mm in diameter, to form a thin cylindrical layer with thickness of 0.5 mm. The temperature of the samples, controlled by thermocouple was maintained during the experiment at $12^{o}$ C.

Agglutinates of as-produced so-called 'ultradisperse diamond' powders were manufactured by detonation synthesis by Gansu Lingyun Nano-Material Co., Ltd., Lanzhou, China. The powders are complex agglomerates up to 20-30 nm in size, consisting of abnormally tight core agglutinates of 60-200 nm in size. This detonation product was disintegrated into primary particles in a vertical-type stirred-media milling machine, designed and manufactured by Kotobuki Industries Co. Ltd., Tokyo, with zirconia beads of 30 nm, manufactured by Toso Co. Ltd., Tokyo. Details of milling and purification operations were described elsewhere [4,5]. In the present study, aqueous colloidal solutions containing about 80 mg/ml of the primary particles having average diameter of 8 nm were used. The colloid, colored intense black believed to be due to holey giant-fullerenic carbon layers on the particle surface, is indefinitely stable in this concentration range. In order to distinguish our dispersed primary particles of detonation nanodiamond from the previously circulated 'ultradisperse diamond' of unspecified particle-size distribution, we call below the former as single-nano buckydiamond (SNBD, further shortly ND) particles.

## III. RESULTS AND DISCUSSION

During preliminary data processing the primary time-of-flight neutron scattering spectra were transformed into the scattering law in the $Q$ - $\varepsilon$ representation (ε - neutron energy transfer under scattering). With the help of program code SLOWN [6] the multiple scattering and the inelastic scattering components of neutron spectra were estimated and removed, and then the quasielastic components were extracted from common neutron scattered spectra in the form of QENS law $S_{QE}(Q,\varepsilon)$ (fig. 1 and 2).

As the first step, quasielastic scattering on BW was analyzed. We did it in detail to get the possibility for the verification of data processing applied and considering BW as the reference system. The QENS law, incorporating effects from the translational and rotational diffusion of water molecules under assumption of their relative independence, can be written as [7-9]:

$$S_{QE}(Q,\varepsilon) = \int S_T(Q,\varepsilon')S_R(Q,\varepsilon-\varepsilon')d\varepsilon' \qquad (1)$$

where $\varepsilon = E-E_o$ and $Q = k-k_o$, are changes in energy and wave vectors of neutron in the scattering event, respectively, $S_T(Q, \varepsilon)$ is the translation component of QENS law and

$$S_R(Q,\varepsilon) = \sum_{l=0}^{\infty}(2l+1)j_l^2(Qa)S_l(\varepsilon) \tag{2}$$

is its rotation component, $j_l(Qa)$ - spherical Bessel function, a is the distance between the molecular centre of gravity and scattering centre for proton. For small values of $Qa$ the sequence of equation (2) can be cut at the third term, and $S_{QE}$ takes the form:

$$S_{QE}(Q,\varepsilon) = j_0^2(Qa)S_T(Q,\varepsilon) + 3j_1^2(Qa)\int S_T(Q,\varepsilon')S_1(\varepsilon-\varepsilon')d\varepsilon' + 5j_2^2(Qa)\int S_T(Q,\varepsilon')S_2(\varepsilon-\varepsilon')d\varepsilon'$$

Suppose the translational and rotational components of QENS law to have lorentzian dependence upon $\varepsilon$, the experimental QENS law looks ultimately:

$$S_{QE}^{EXP}(Q,\varepsilon) = \{\frac{1}{\pi} j_0^2(Qa)\left[\frac{\Delta E_T(Q,\varepsilon)}{\varepsilon^2 + \Delta E_T^2(Q,\varepsilon)}\right] +$$
$$+3j_1^2(Qa)\frac{1}{\pi}\left[\frac{\Delta E_T(Q,\varepsilon)+\Delta E_{R1}(\varepsilon)}{\varepsilon^2 + (\Delta E_T(Q,\varepsilon)+\Delta E_{R1}(\varepsilon))^2}\right] + 5j_2^2(Qa)\frac{1}{\pi}\left[\frac{\Delta E_T(Q,\varepsilon)+\Delta E_{R2}(\varepsilon)}{\varepsilon^2 + (\Delta E_T(Q,\varepsilon)+\Delta E_{R2}(\varepsilon))^2}\right]\} \otimes R(Q,\varepsilon) \tag{3}$$

where $\Delta E_T(Q,\varepsilon)$ is halfwidth at halfmaximum (HWHM) of the translational components, $\Delta E_{R1}(\varepsilon)$ – HWHM of the first rotational component, and $\Delta E_{R2}(\varepsilon)$– HWHM of the second rotational component (under an assumption that the reorientation of the molecule proceeds according to a simple diffusion model $\Delta E_{R2}(\varepsilon) = 3\,\Delta E_{R1}(\varepsilon)$); $R(Q, \varepsilon)$ - resolution function of spectrometer for initial energy, measured with a standard vanadium sample. An example of the experimental QENS law decomposition into the translational and rotational components as well as the degree of concordance of model and experimental curves are demonstrated in fig. 3.

The translational component of QENS law $S_T(Q, \varepsilon)$ was described using the model of mixed diffusion [10]. It means, the diffusion mobility of water molecule includes two mechanisms: the jump diffusion (with the parameter $\tau_o$ - the residence time of molecule at the temporary equilibrium position) and the continuous diffusion of molecule (with the parameter $D_0$ – the coefficient of continuous common diffusion of molecule together with the nearest surrounding). In the frame of this model the quasielastic peak proves to be lorentzian with the FWHM:

$$\Delta E(Q^2) = FWHM = 2\frac{\hbar}{\tau_0}\left[1 + D_0\tau_0 Q^2 - \frac{\exp(-<u^2>Q^2)}{1+(D-D_0)\tau_0 Q^2}\right] \tag{4}$$

where $D$ – is the total self-diffusion coefficient in water.

The FWHMs of the translational and rotational components of QENS law (3), and the parameters of the model (4) for BW, that were inferred from the optimal description of the experimental data by the equation (4), are shown in fig. 4. As a whole, the results we obtained in BW, for both translational and rotational diffusion of water molecule were found to be similar to the generally accepted values [8].

Before to discuss the results for (DW) it should be noted that the assumption concerning the difference in spectra scattered by BW and DW (fig. 2) can not be assigned to the own carbon particles scattering. Carbon possesses only the coherent scattering, which can be observed for $Q$ $> 3$ Å$^{-1}$ (crystal lattice spacing for diamond $d = 3.567$ Å [11]), that is beyond of Q-limits, covered in our experiment.

Before the analysis of DW spectra the remark should be made. According to the point of view, based on the theoretical calculations [12-14] and molecular dynamics (MD) simulations [15-19], a CNP being inserted into water, causes the substantial changes in the structure and microdynamic properties of surrounding waters, which extends only over the nearest 2- 3 layers of water molecules (no far than ~ 10 Å). Keeping this in mind we assumed DW to be a two-component system comprising of BW and a small additive belonging to the nearest surrounding of ND and subjected to its influence (HW). This circumstance was taken into account by two ways. First, by the introduction into the expression (3) of the complementary term in the single lorentzian form. The relative fractions of the main (BW) and extra (HW) components were connected through relation $n_{BW} + n_{HW} = 1$. The degree of accordance for the experimental curves and the description of DW by two-component model mentioned above, is demonstrated in fig. 5.

The second way of the estimation of the relative HW fraction consisted in the direct subtraction of DW and BW quasielastic spectra, normalized in the region of the wings, where the effects of HW presence are presumably absent (energy transfer $\varepsilon = \pm (0.5 - 1.4)$ $meV$, see fig. 6). The results of such an analysis are also presented in fig. 7. Due to remarkable uncertainties of these results the both sets of data can be estimated as coinciding ones and giving the average relative fraction of HW as $n_{HW} \sim (3.3 \pm 0.3)$ %. This number of molecules corresponds to nearly three water layers, surrounding ND. Keeping in mind the concentration of ND in the dispersion under investigation, we estimated the quantity of HW per 1 g of ND as ~ 0.3 g HW/g ND.

As the next step, we have undertaken the attempt to analyze the quasielastic peaks of. HD, whose FWHM's are shown in fig. 8. In so doing, we have kept in mind the smallness of the HW relative fraction and an expected considerable uncertainties of the parameters we tried to get. For the analysis of quasielastic HW peaks two ways were used. As the first one, for the description of the curve of fig. 8 we applied the model of mixed diffusion (4). Fig. 9 shows the results of such an analysis.

It is seen, the diffusion mobility of HW molecules appeared to be remarkably slowed down comparing to BW, due to the reorientation mobility $((\Delta E_{rot1})_{BW}/(\Delta E_{rot1})_{HW} \sim 20)$, as well as translation diffusion (the residence time of HW molecule increases approximately 4 times). The comparison of these results with the experimental data for supercooled water [8] leads to the conclusion, that the diffusion mobility of HW in our conditions looks similarly to that of supercooled BW at the temperature about – (15 – 17)$^{o}$ C.

The direct comparison of our results with MD data for carbon surface (graphite, graphen ) is faced with difficulties due to strong spatial asymmetry of selfdiffusion coefficient, obtained in such conditions by MD simulations, which our experiment is enable to distinguish.

At the same time there exist other ways for the analysis of the QENS results, elaborated on the basis of the mode coupling theory (MCT) [20] in application to liquids in confined geometry or supercooled conditions. In particular, the model of relaxing cage is widely used [21-23]. According to this model, the diffusion act in the liquid can be realized only under conditions when the structural decay of surrounding for a given particle takes place. These processes are called as α – relaxation (slow relaxation) [24, 25]. In the frame of this approach the intermediate scattering function (ISF) is presented in the form of "stretched exponent", which takes into account the possible non-exponential decay of the ISF:

$$I(Q,t) = \exp\left[-(t/\tau_W)^{\beta}\right] \quad (5)$$

Such a presentation of ISF results in the nonlorentzian form of the QENS law (or dynamical scattering function, DSF):

$$S_{QE}(Q,\varepsilon) = \frac{1}{\pi\hbar}\int_0^{\infty} \exp\left[-(t/\tau_W)^{\beta}\right] * \cos(\varepsilon * t/\hbar)dt \qquad (6)$$

where $\tau_W$ – time of $\alpha$ – relaxation (the decay time of the nearest surroundings for a given particle), $\beta$ – a form parameter (for $\beta = 1$, $S_{QE}(Q,\varepsilon)$ turns to be lorentzian). These parameters as well as the average relaxation time

$$<\tau_W> = \int_0^{\infty} dt * \exp\left[-(t/\tau_W)^{\beta}\right] = \frac{\tau_W}{\beta}\Gamma(\frac{1}{\beta}) \qquad (7)$$

are functions of $Q$, and $<\tau_W>(Q)$ has the form

$$<\tau_W> = \tau_0 Q^{-\gamma} \qquad (8)$$

When dealing with the simple continuous diffusion, $\gamma = 2$, but under supercooling or confined geometry conditions it appears, that $\gamma < 2$. The methods of the separation and accounting for the translation and rotation effects in the diffusion mobility of water in the frame of this model are carefully discussed in [26]

Our experimental results, treated on the grounds of the exp. (5-8), are shown in fig. 10. The relaxation parameters we get for BW (lower part of fig. 10) are close to those known at the corresponding temperature [27]. The average relaxation time of HW (upper part of fig. 10) far exceeds that of BW. It confirms above conclusion that the diffusion processes in the nearest vicinity of ND are markedly slowed down. According to [27], the relaxation characteristics of HW we get, roughly agree with those of BW supercooled at ~ - (15 – 20)$^o$ C.

A summary of the diffusion parameters for BW and HW, we obtained by analyzing the experimental QENS data on ND aqueous dispersion, is shown in the Table 1.

Table 1. Diffusion characteristics of bulk and hydration water

| Diffusion characteristics | Bulk water (BW) | Hydration water (HW) |
|---|---|---|
| D, $10^{-5}$ $cm^2/s$ | 1.54 ±0.06 | 1.05 ± 0.1 |
| $D_o$, $10^{-5} cm^2/s$ | 0.04±0.01 | 0 |
| $\tau_o$, ps | 2.22 ±0.04 | 8.7 ± 0.7 |
| $\Delta E_{rot1}$, meV | 0.43± 0,03 | ~ 0.02 |
| $\tau_{rot1}$, * ps | 3± 0,02 | ~ 65 |
| $<\tau_w>$, **ps | ~ 8 | ~ 30 |

* $\tau_{rot1} = 2\hbar / \Delta E_{rot1}$.

***for* $Q \sim 1 Å^{-1}$.

## IV. CONCLUSION

The inelastic neutron scattering experiment was performed on BW and DW - concentrated aqueous dispersion of ND. The quasielastic component was inferred from the experimental double-differential scattering cross sections and analyzed. The diffusion characteristics obtained for BW, as reference system, appeared to be about the same as reported in the literature.

The QENS spectra of DW were analyzed by two ways. In the first one DW was described as the two-component system, consisted of BW and a small additive having presumably the lorentzian form. This additive is called as hydration water (HW), located in the nearest surrounding of ND's. The second approach included the direct subtraction of DW and BW spectra, properly normalized in the regions of their wings, where the influence of HB can be neglected. The difference of this subtraction was considered as fraction of HW, existing in DW. The relative fractions of HB, extracted for both cases, in the limits of errors are near each other and equal about 3 %. This corresponds to about thee layers, surrounding ND particle.

As to diffusion properties of HW, which estimates are of the conditional character, they demonstrate the remarkable deterioration of molecular mobility compared to BW: the rotational mobility is decreased in ~ 20 times, the residence time and the average relaxation time of molecules are increased in ~ 4 times. The comparison of our HW to supercooled BW in the diffusion mobility leads to the conclusion, that the state of our HW corresponds to the latter at the temperature about ~ – $(15\text{-}20)^{o}$ C.

It should be mentioned that the diffusion characteristics for HW we present, are averaged ones over about three molecular layers adjacent to the surface of hydrophobic particle. In the very first adjacent layer the hydration effects discussed above are expected to manifest themselves more distinctly.

**Acknowledgements**

Authors gratefully acknowledge the financial support of ISTC (project # 2769).

FIGURE CAPTIONS

**Fig.1** QENS spectra of BW for some values of wavevector transfer Q. Resolution HWHM = 0.07 meV.

**Fig.2** Typical QENS spectra for DW and BW for Q = 0.98, 1.57 и 2.13Å$^{-1}$, normalized by wings in the region ε = ± ( 0.5 – 1.4) meV. Full symbols – DW; open symbols – BW.

**Fig.3** Typical decomposition of QENS peak for BW into translational and rotational components.

**Fig.4** FWHM-s of translational and rotational components for BW QENS peaks. Full curve – description by model (4).

**Fig.5** Typical decomposition of QENS peak for DW into translational and rotational components, corresponding to BW, and additive hydration fraction.

**Fig.6** Typical QENS spectra of HW (full symbols), obtained by direct subtraction of DW (open symbols) and BW (continuous lines) QENS spectra, normalized by wings in the region ε = ± ( 0.5 – 1.4) meV.

**Fig.7** Relative fraction of HW component in DW ($< n > = (3.3 \pm 0.3)$ %) obtained by exp. (3) with complementary lorentzian (full symbols) and by direct subtraction of DW and BW QENS spectra (open symbols).

**Fig.8** FWHM of QENS spectra for HW. Full line – approximation by model (4).

**Fig.9** Decomposition of FWHM for HW into translational and rotational components. Solid curve – description by model (4), dashed and thin lines - supercooled water, T= - 15 $^{o}$C and -17 $^{o}$C correspondingly [8]. Shaded area depicts the errors of FWHM.

**Fig.10** Upper part: shaded area – Q - dependence of the experimental average α- relaxation time in HW; full line - its approximation by $<\tau_W>(Q) = 31*Q^{-1.7}$ ($\gamma = 1.7\pm0.1$ is estimated for 0.5 $A^{o-1} < Q < 1.4A^{o-1}$ region); dashed line – MD data [27] for – 20$^{o}$ C. Shaded area depicts the errors of experimental $<\tau_W>$.

Lower part: points - Q - dependence of experimental average α-relaxation time in BW; full line – its approximation by $<\tau_W>(Q) = 9.0*Q^{-2.02}$ ; dashed line – MD data [27] for T=284K.

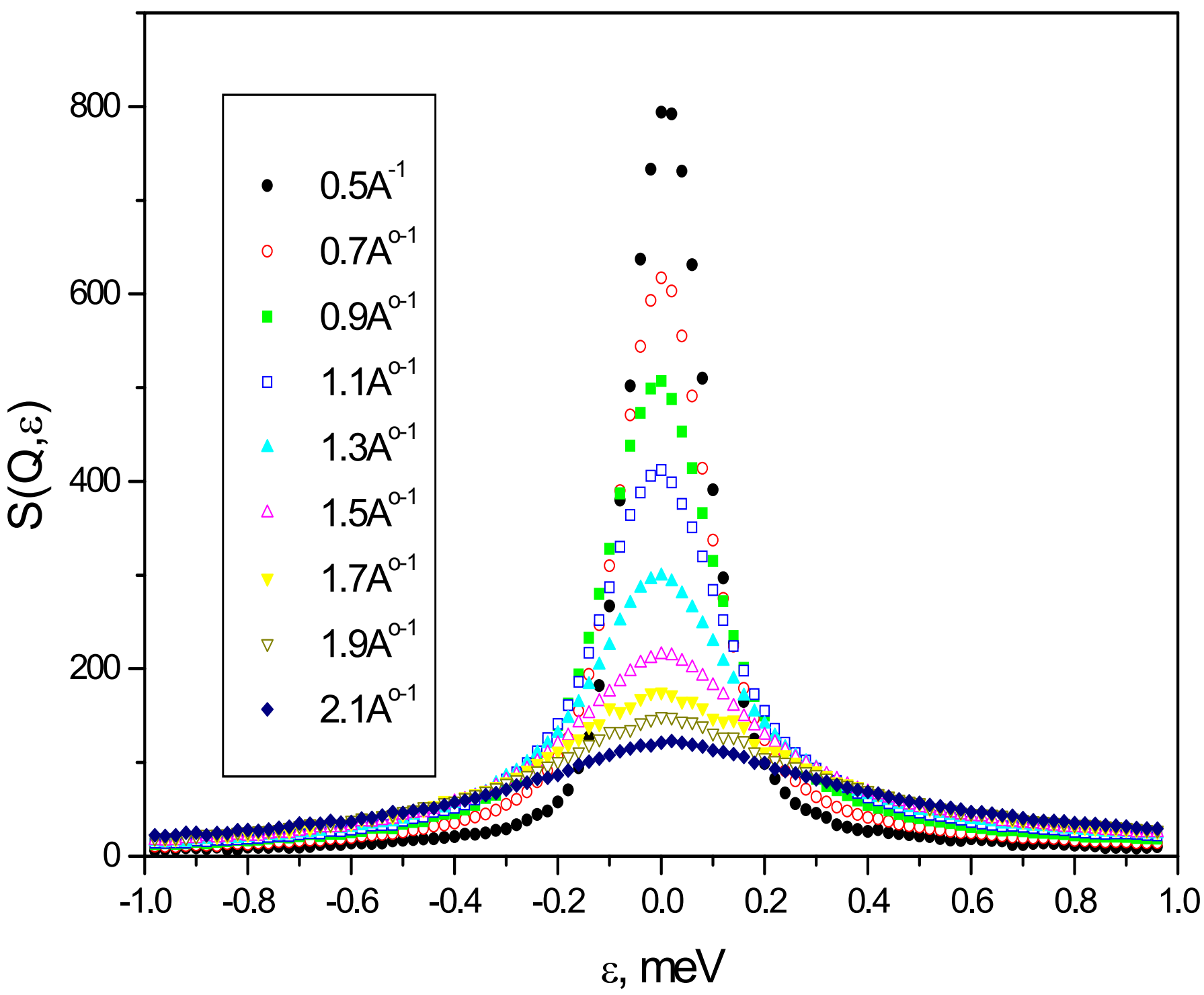

**Fig.1**

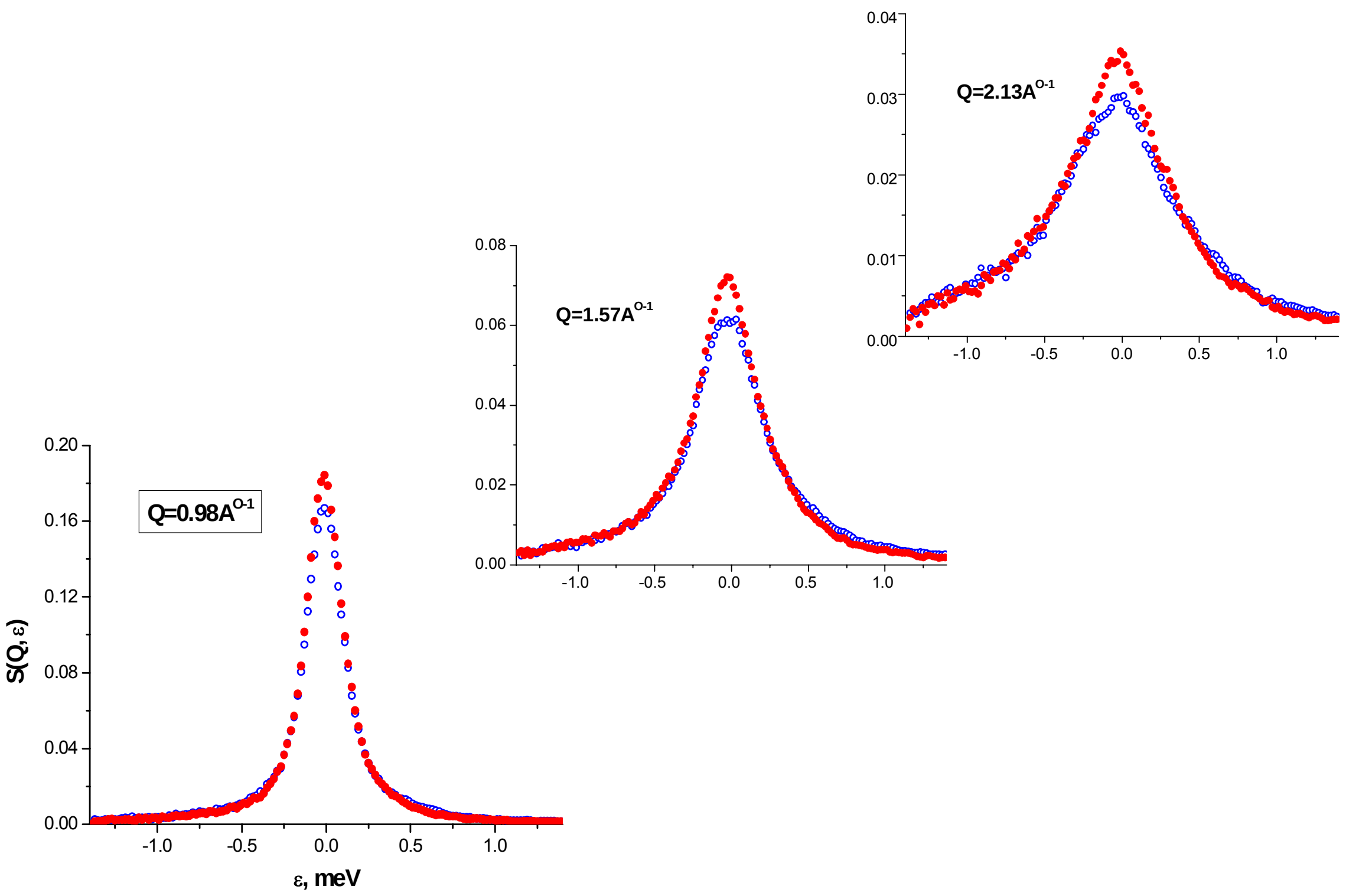

**Fig. 2**

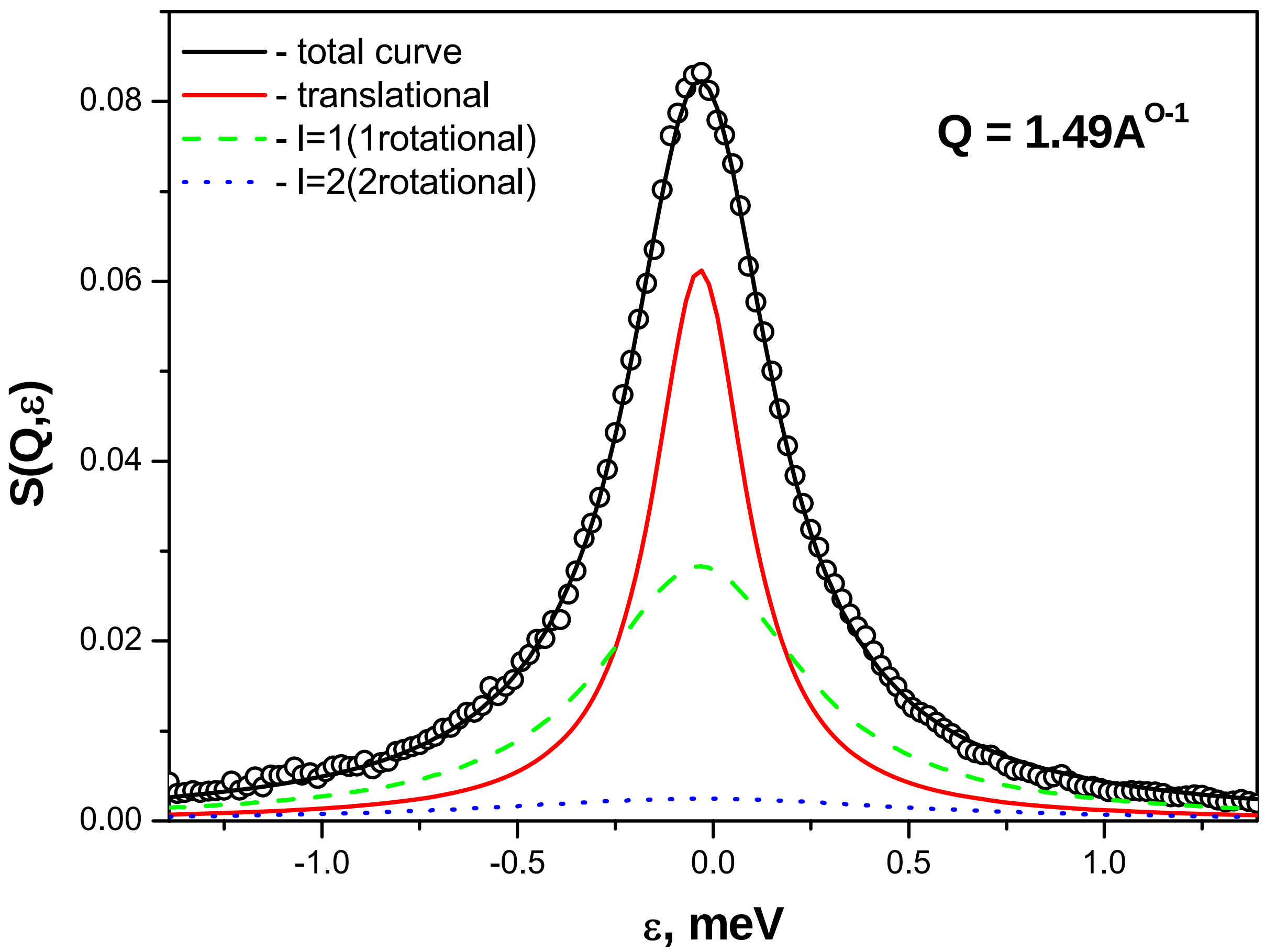

**Fig.3**

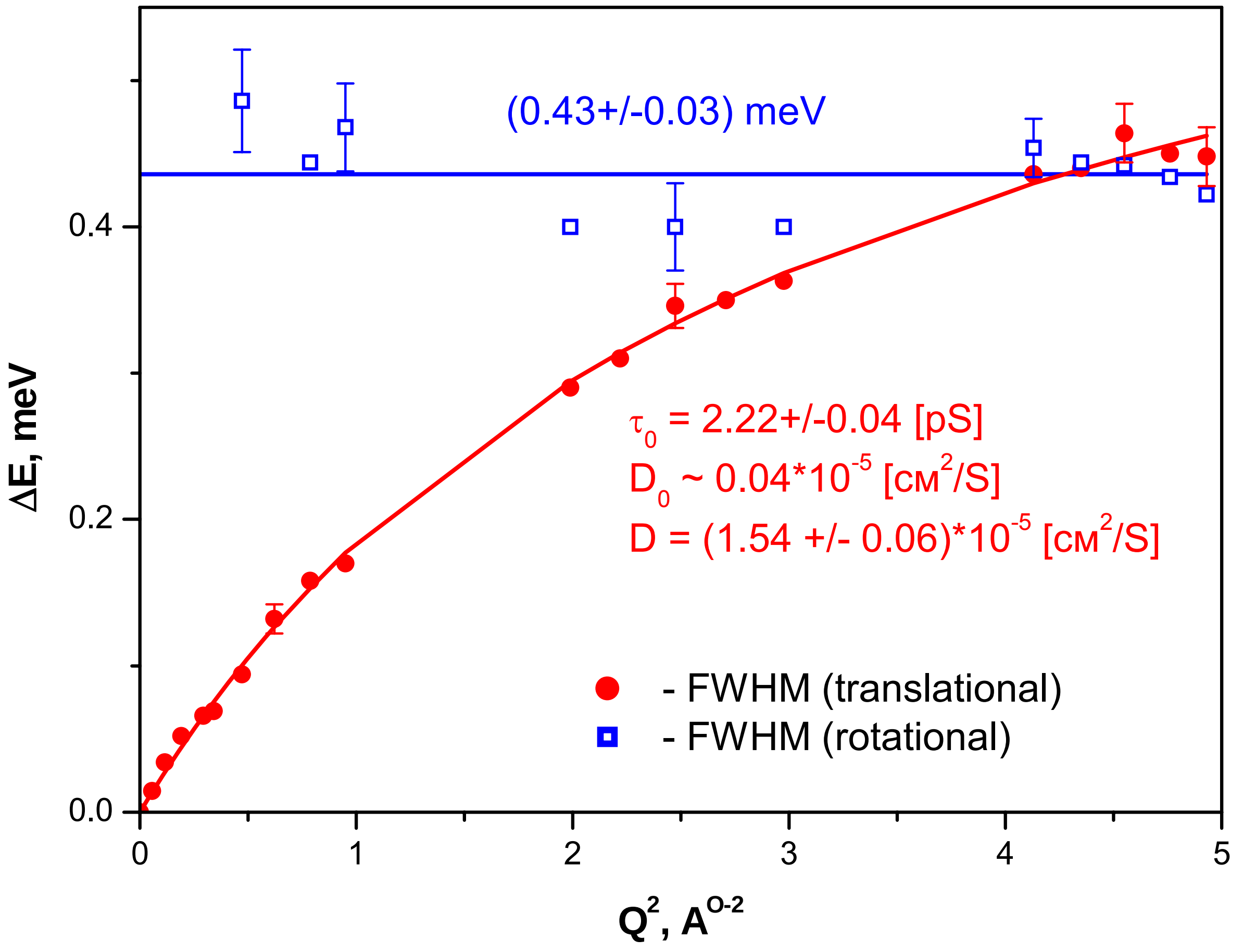

**Fig.4**

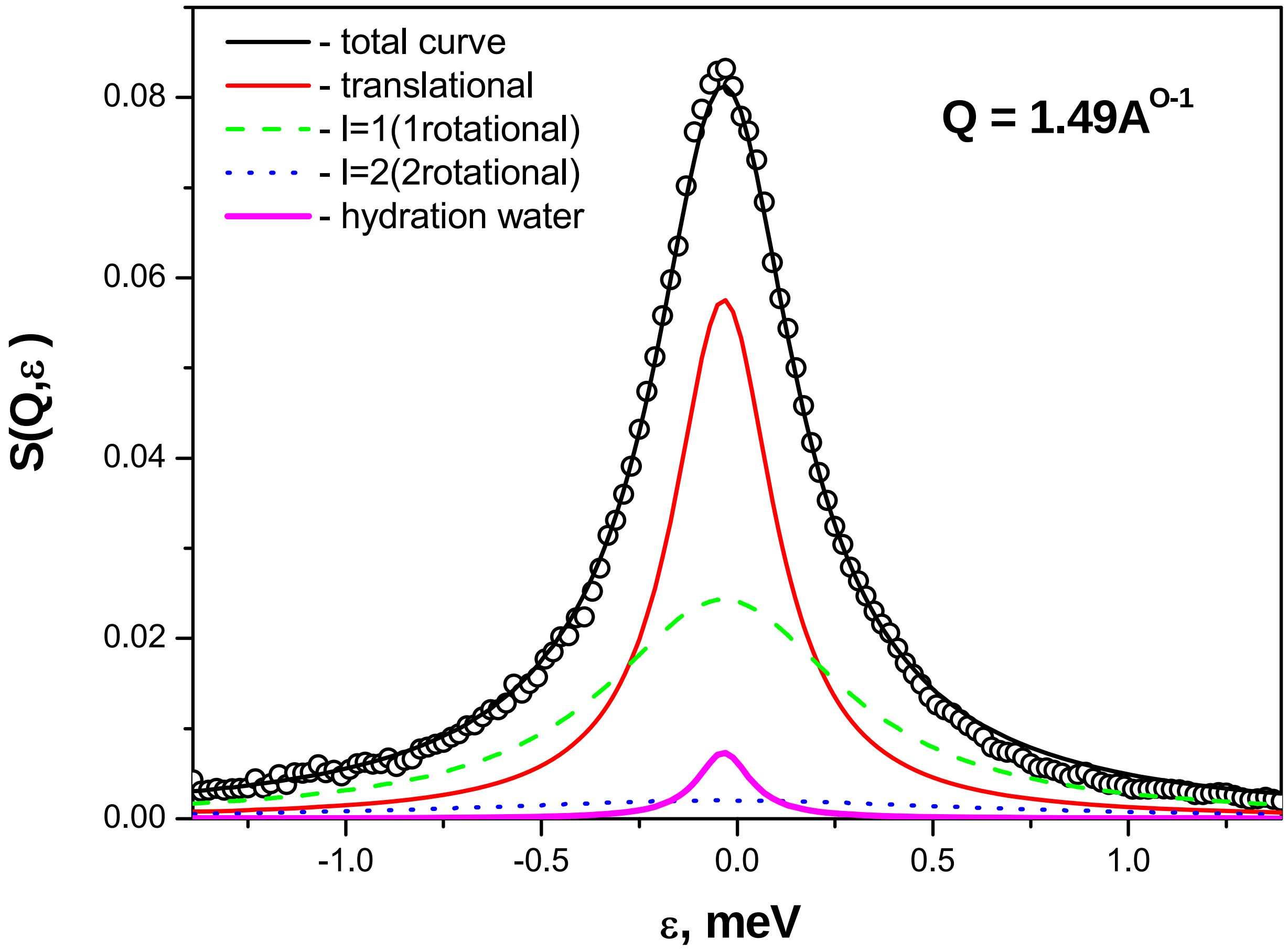

**Fig.5**

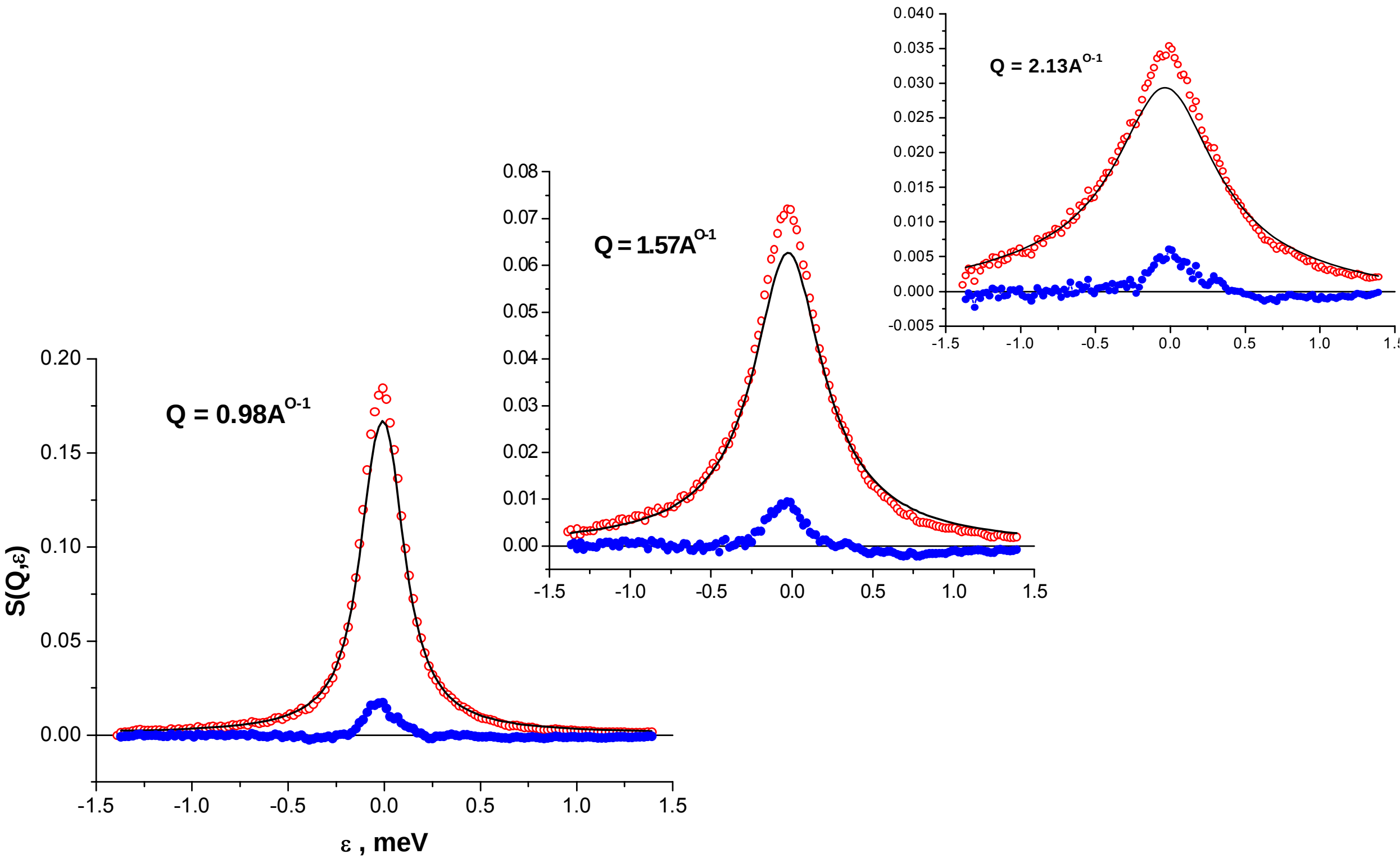

**Fig.6**

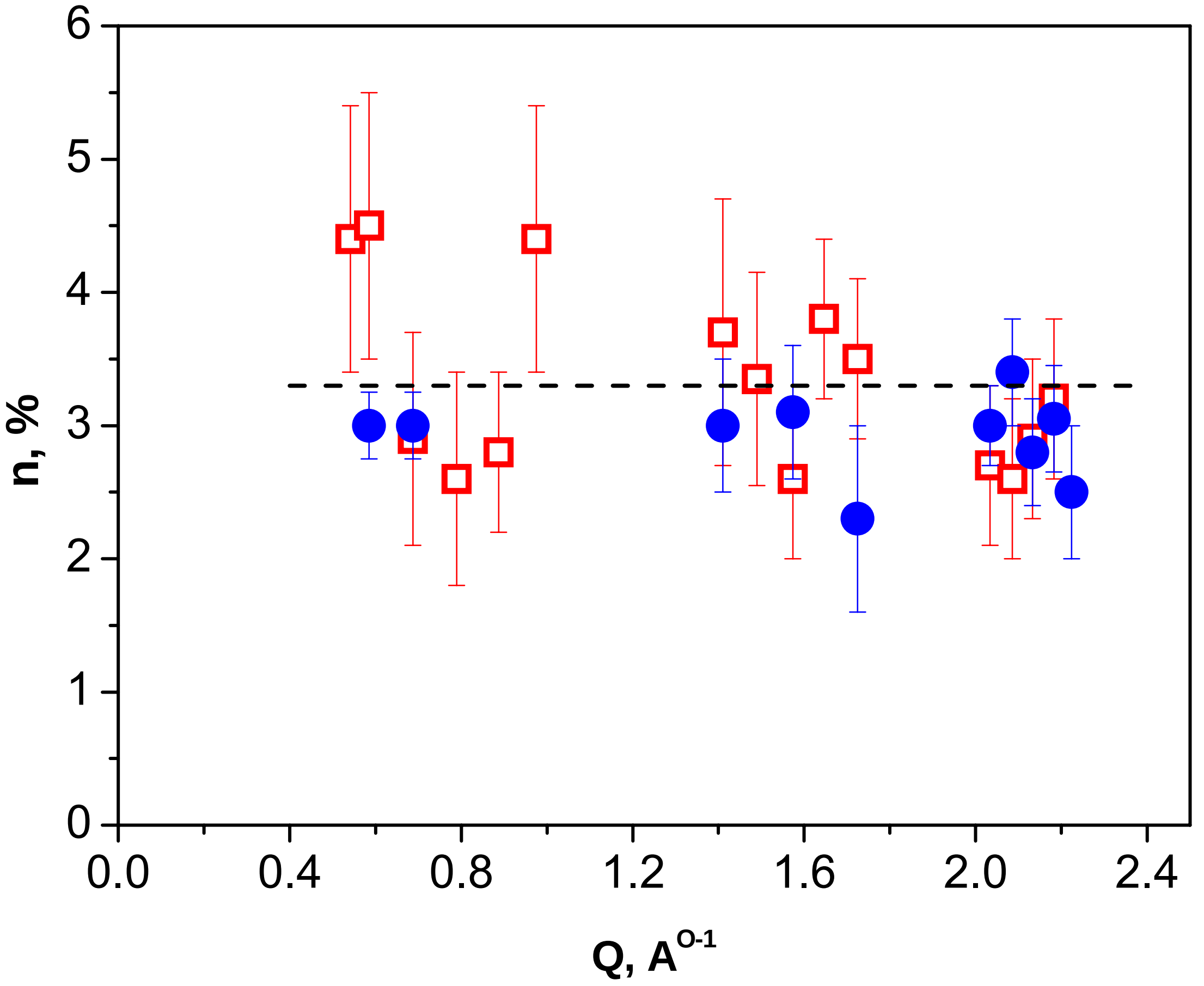

**Fig.7**

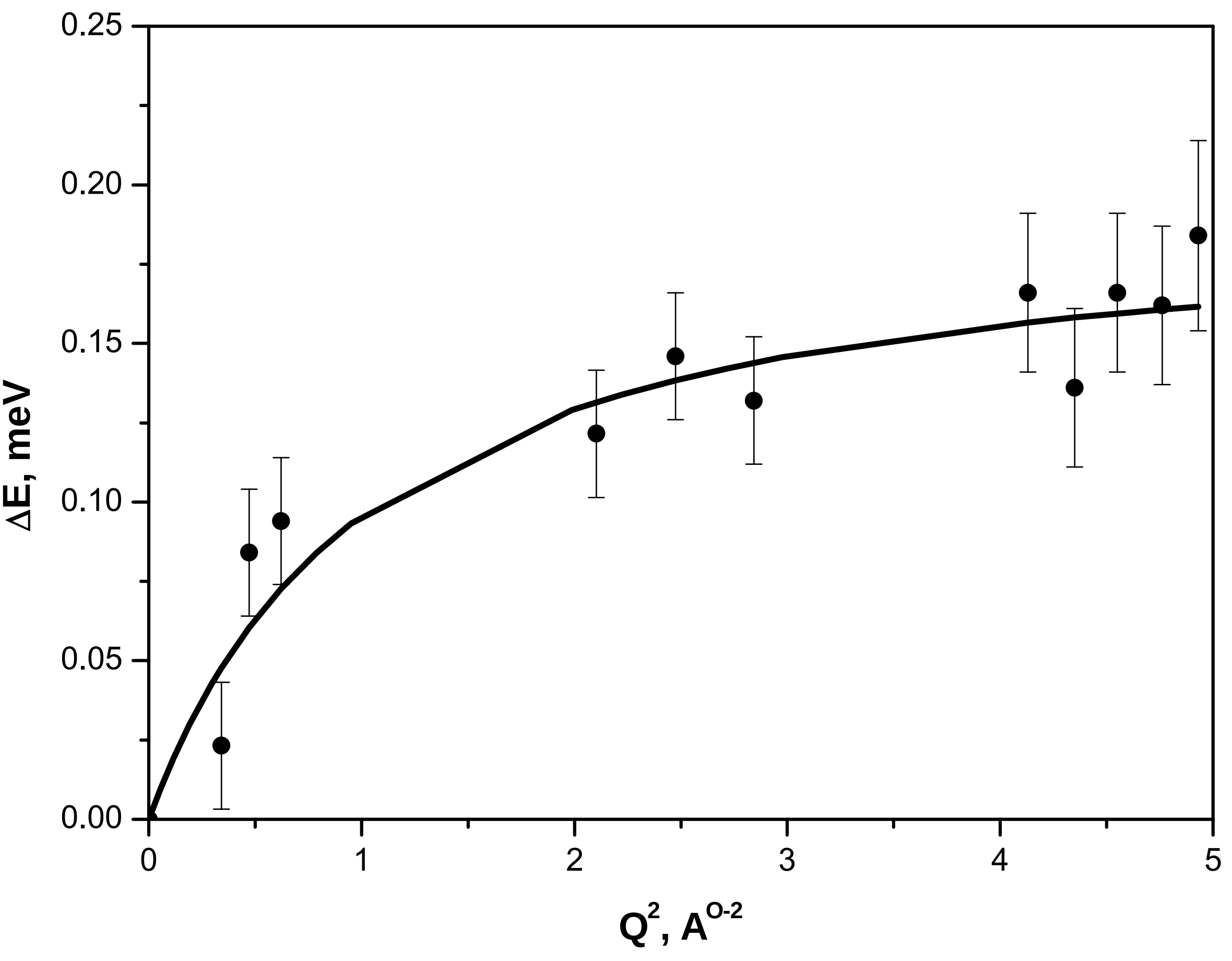

**Fig.8**

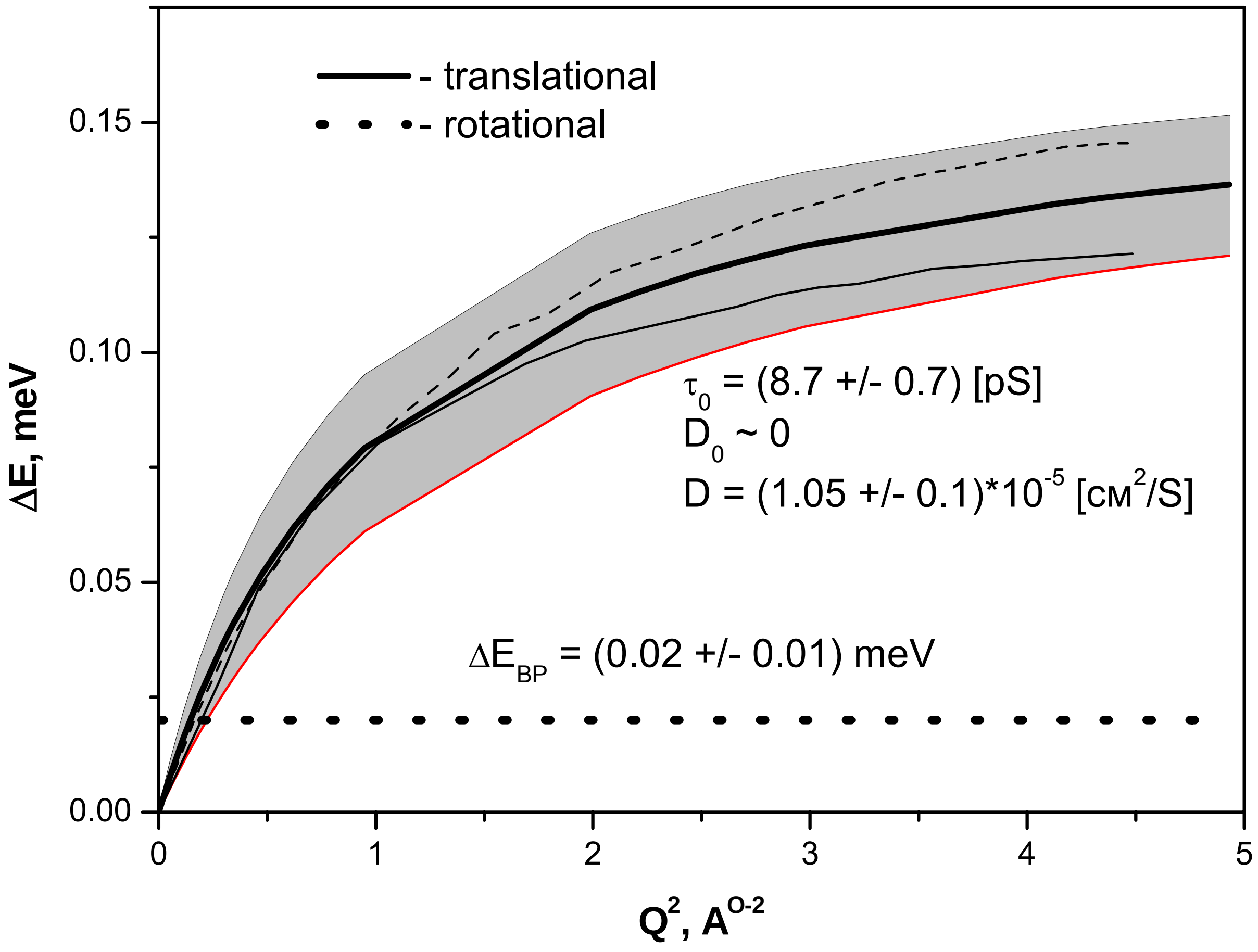

**Fig.9**

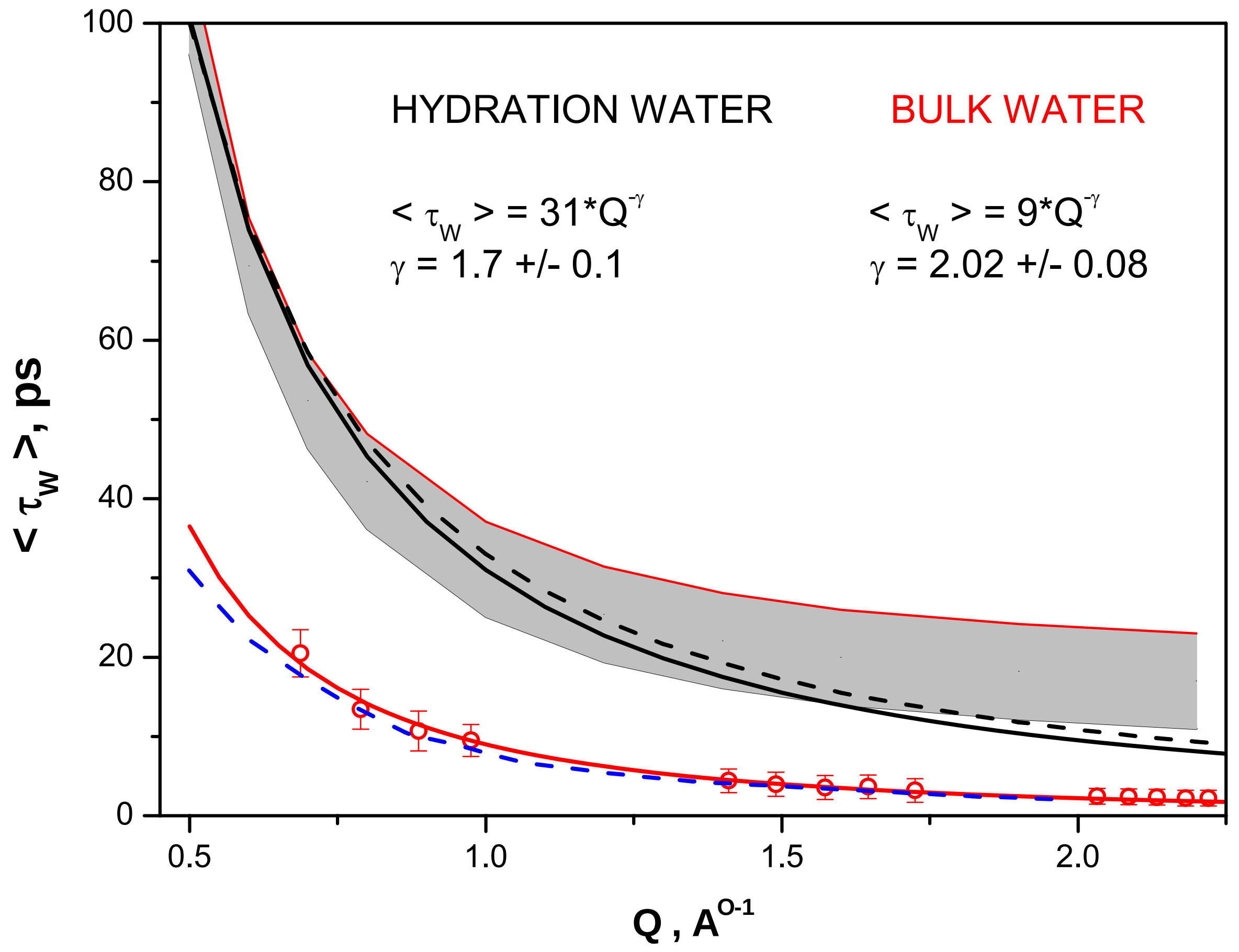

HYDRATION WATER
BULK WATER
< τW > = 31*Q-γ
γ = 1.7 +/- 0.1
< τW > = 9*Q-γ
γ = 2.02 +/- 0.08
< τW >, ps
Q , AO-1
100
80
60
40
20
0
0.5
1.0
1.5
2.0

**Fig.10**